\documentclass[a4paper]{spie}
\usepackage{graphicx}
\title{Laboratory Testing the Layer Oriented Wavefront Sensor for the Multiconjugate Adaptive optics Demonstrator}
\author{Carmelo Arcidiacono\supit{a}, {Matteo Lombini\supit{a,b}},Emiliano Diolaiti\supit{b}, Jacopo Farinato\supit{b}, Roberto 
Ragazzoni\supit{c}\\
\supit{a}INAF - Osservatorio Astrofisico di Arcetri - Largo E.Fermi 5, 
I-50125 Firenze Italy; \\
\supit{b}INAF - Osservatorio Astronomico di Bologna - Via Ranzani 1, 
I-40127 Bologna Italy; \\
\supit{c}INAF- Osservatorio Astronomico di Padova - Vicolo 
dell'Osservatorio 2, I-35122 Padova Italy;\\
}
\authorinfo{Further author information: (Send correspondence to Carmelo Arcidiacono)\\E-mail: carmelo@arcetri.astro.it, Telephone: +39 055 2752 293\\              }

\begin{document}

\maketitle

\begin{abstract}
The Multiconjugate Adaptive optics Demonstrator (MAD) for ESO-Very Large 
Telescopes (VLT) will demonstrate on sky the MultiConjugate Adaptive 
Optics (MCAO) technique. In this paper the laboratory tests relative to 
the first preliminary acceptance in Europe of the Layer Oriented (LO) 
Wavefront Sensor (WFS) for MAD will be described: the capabilities of 
the LO approach have been checked and the ability of the WFS to measure 
phase screens positioned at different altitudes has been experimented. 
The LO WFS was opto-mechanically integrated and aligned in INAF - 
Astrophysical Observatory of Arcetri before the delivering to ESO 
(Garching) to be installed on the final optical bench. 
The LO WFS looks for up to 8 reference stars on a 2arcmin Field of View 
and up to 8 pyramids can be positioned where the focal spot images of 
the reference stars form, splitting the light in four beams. Then two 
objectives conjugated at different altitudes simultaneously produce a 
quadruple pupil image of each reference star.

An optical bench setup and transparent plastic screens have been used to 
simulate telescope and static atmospheric layers at different altitudes 
and a set of optical fibers as (white) light source.

The plastic screens set has been characterized using an inteferometer 
and the wave-front measurements compared to the LO WFS ones have shown 
correlation up to $\sim$95\%.
\end{abstract}


\section{INTRODUCTION}
The Multiconjugate Adaptive optics Demonstrator (MAD) will be the first 
Multi Conjugate Adaptive Optics (MCAO) instrument working on sky. It 
will be mounted aboard the Nasmyth platform of the UT-3 Very Large 
Telescopes (VLT). In this paper the preliminary results of the 
laboratory tests performed on the Layer Oriented (LO) Wavefront Sensor 
(WFS) for MAD will be presented: the Wave-Front (WF) reconstruction 
capabilities of the LO approach have been checked and the ability of the 
WFS to measure phase screens positioned at different altitudes have been 
experimented. The Layer Oriented (LO) approach\cite{LO2} had been tested 
in laboratory so far using a mini-prototype of the MAD LO {WFS\cite{2004SPIE.5382..578F}}. 
The LO WFS for MAD has been aligned and tested at the INAF 
(Astrophysical Observatory of Arcetri) where it was mounted on an 
optical bench and making part of an optical set-up intended to mimic the 
F/20 MAD focus at the LOWFS base. Hereafter this optical setup is called 
``\,telescope simulator". It simulates infinitely far sources over the 
2arcmin technical FoV of the WFS, the telescope pupil and a turbulent 
static-multi-layers atmosphere.

The LO WFS has a technical Field of View of 2arcmin and up to 8 pyramids 
can be positioned where the focal spot images of the reference stars 
form, splitting the light in four beams. Then two objectives conjugated 
at different altitudes simultaneously produce a quadruple pupil image of 
each reference star.

Two optical relays compose the so called Ground and High layer WFS, they 
can be optically conjugated to altitudes between 0 and 18 km by simply 
positioning the CCD using two as many linear stages, controlled via 
software by the Instrument Control Software (ICS). For each reference 
star four geometrically identical pupil images are imaged on the two 
CCDs, and the intensity variations due to different WF distortions are 
measured simultaneously for all the references. Finally the local 
tip-tilt can be computed by simply measuring the illumination difference 
between same portions of different pupils. In the experimental setup 
installed in Arcetri the metapupils (pupil plane projection of the Field 
of View at the conjugation altitude) are imaged on a couple of temporary 
CCDs (Electrim 1000N) because the final two MARCONI CCD39s were not 
available. Electrim 1000N uses 4.75$\times$3.66 mm sensing area divided in 
652$\times$494 pixels. Because the correspondent pixel size is 7.4$\mu$m, finer than 
the CCD39s 24$\mu$m, it was possible to achieve a very high precision 
(better than a tenth of sub-aperture) in the alignment phase. In order 
to shrink the dimension of the pupils on the CCD the F number from the 
initial F/20 have been enlarged to F/300 using a simple optical relay\cite{tubetti} 
composed by two small lenses placed on the same mounting of the 
pyramids in order to be placed in correspondence of reference star 
position on the focal plane. In this way the dimension of the spot on 
the pyramid is enlarged of factor 15 while the pupil dimension is shrank 
by the same factor, giving finally the 384m of the pupil dimension. 
These optical relay will be called hereafter ``\,star enlarger". Several 
tests have been performed to check the optical quality, the alignment 
and the WF measurements repeatability. Moreover the LO WFS measurements 
have been calibrated to translate the slope measurements in nm of the WF 
using theoretical formula and ray-tracing simulations considering the 
defocus introduced by moving of fiber light sources toward and forward 
with respect the zero defocus positions along the optical axis 
direction. A set of plastic screens has been characterized using an 
inteferometer and the WF measurements compared to the LO WFS ones 
showing very high correlations (Perfect correlation 1, No correlation 0, 
Anticorrelation -1). Finally two tests are presented: 

1. The verification of the sensor ability to measure phase screen WF at 
the non-conjugation altitudes;

2. The verification of the sensor ability to measure the WF of 
conjugated plane inserting, in different orders, two screens at the two 
conjugation altitudes and one in between.

More and detailed information about the LOWFS for MAD could be found in 
the paper Vernet et al.\cite{vernet2005}

\section{The Telescope Simulator}

\begin{figure}[h]
\centerline{\includegraphics[width=8.75cm,height=6.42cm]{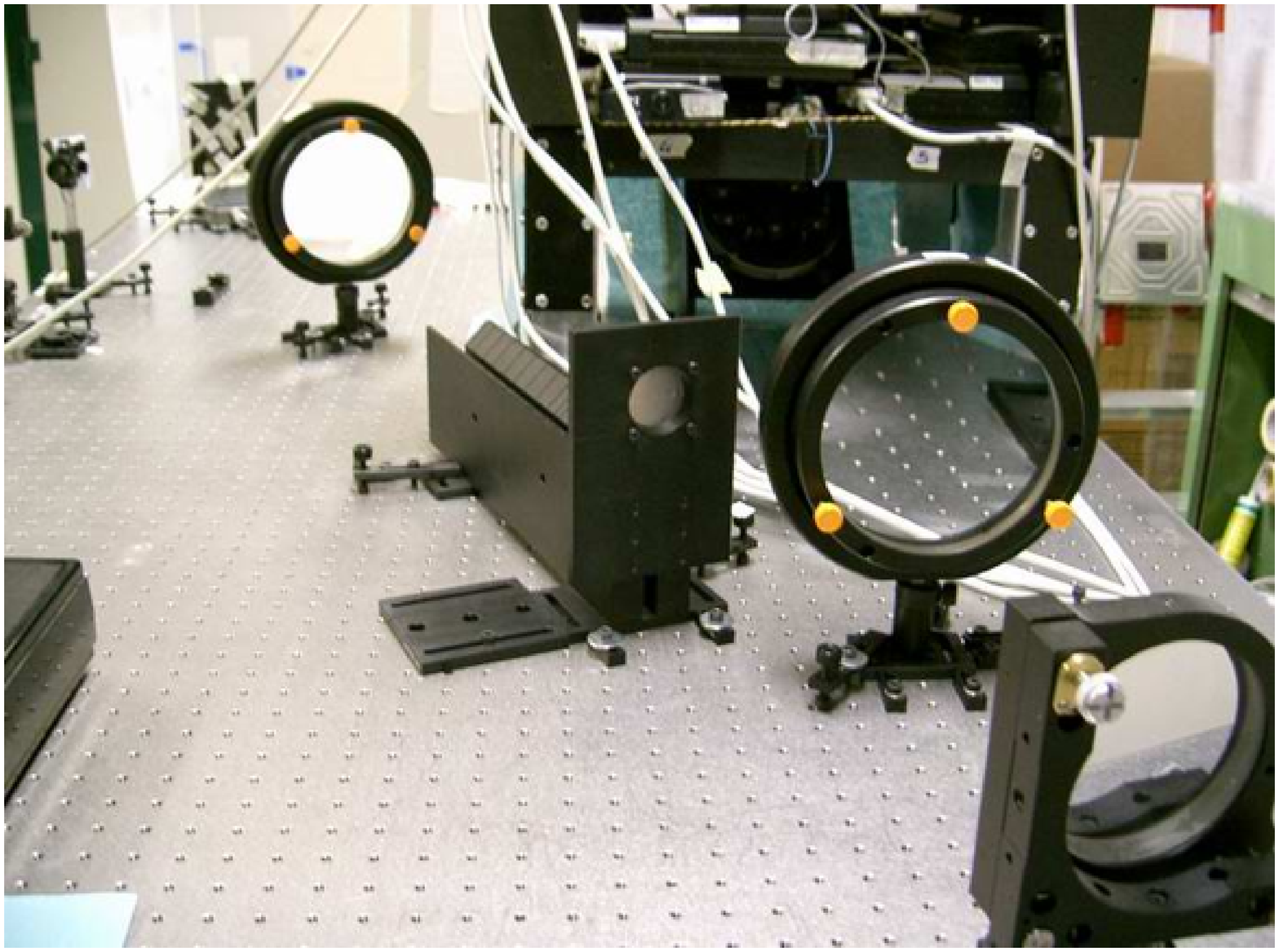}
\includegraphics[width=228pt,height=181.5pt]{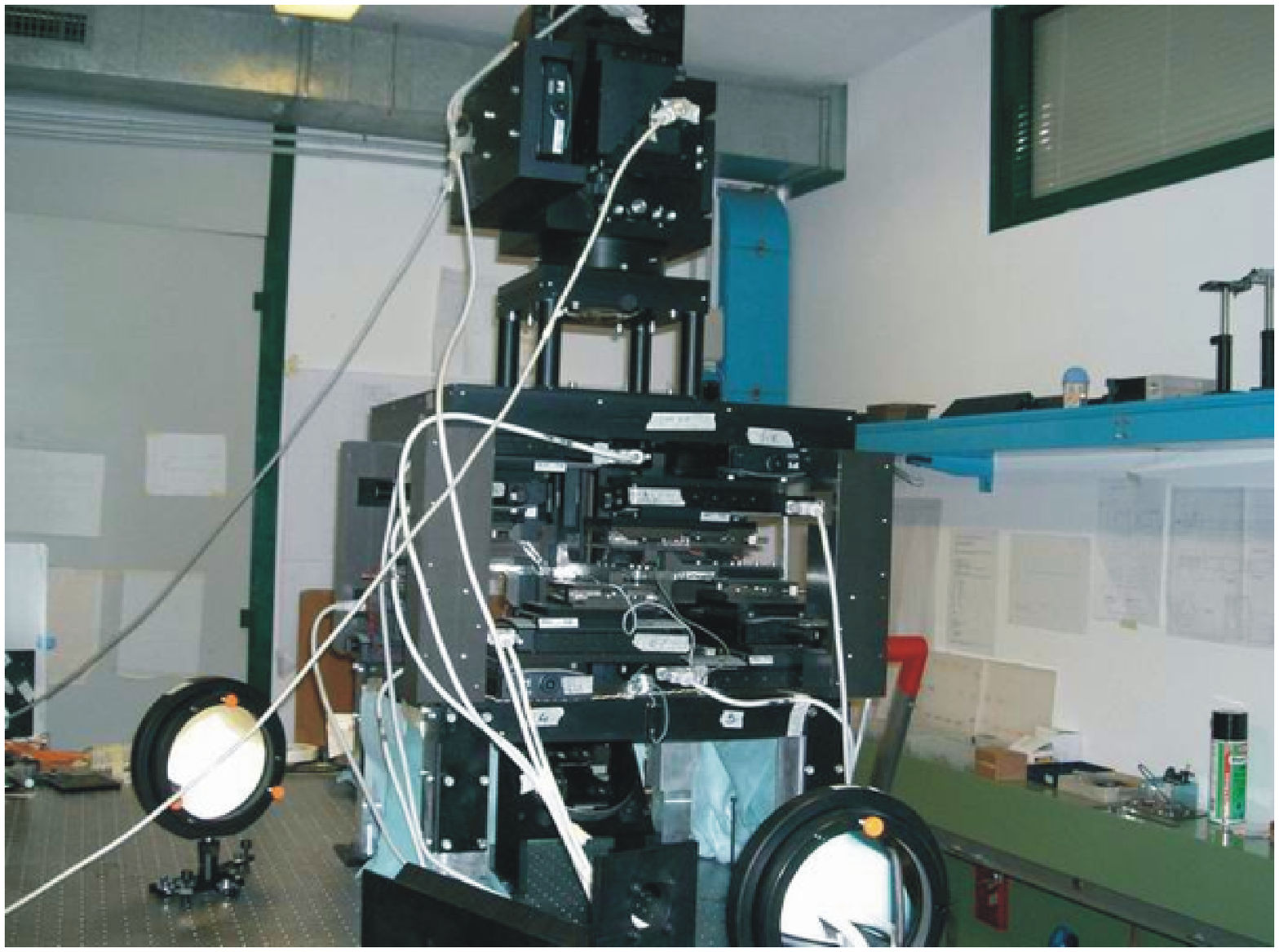}}
\caption{\footnotesize{Left: This picture shows the Telescope Simulator 
optical setup. On a corner of the bench (top-left in this image) the 
plate for fibers positioning. The telescope simulator is composed by two 
identical 150mm lenses spaced by twice their focal length (793mm) and a 
stop pupil in their common foci. A turbulent static screen holder is 
arranged to place them on corresponding position between 0 and 18km from 
the ground pupil. Right: the LOWFS MAD mounted on the bench. The 
telescope simulator is positioned to match its F/20 focal plane with the 
LOWFS entering focal plane. All the system has been aligned using as 
reference the optical axis defined by the common lenses of the pupil 
re-imager. Connected to the wires it is possible to identify the XY 
linear stages that move the pyramids over the 2arcmin FoV. The higher 
wires are connected to the CCDs.}}\label{fig:1}
\end{figure}

The telescope simulator intends to mimic the F/20 MAD focus at the LOWFS 
base positioned that, according to the opto-mechanical drawings, falls 
10 mm above the mechanical base of the system (by drawings 4.35mm before 
the position of the star enlargers). At this plane is projected the 2 
arcmin FoV seen by the sensor. In the alignment optical setup used the 
LOWFS base-plate stands on four legs to leave room below it for a mirror 
at 45 degrees with respect to the optical axis direction to fold this 
inside the LOWFS.

The experimental optical setup has been designed in a way similar to the 
system used to test the LO {prototype\cite{2004SPIE.5382..578F}}. It will allow testing, 
in open loop conditions, the stand-alone LOWFS by using white light 
emitted by a set of fibers and distorted by a set of perturbing screens 
to simulate the atmospheric turbulent layers.

Two identical lenses spaced, by twice their focal length, and a stop 
pupil, in their common foci, compose the telescope simulator. A 
turbulent static screen mounting is arranged to place them on positions 
corresponding between 0 and 18km from the ground pupil. 

Two lenses of 150mm diameter with a focal length of 793mm are used to 
mimic the telecentric beam. The optical axis flies parallel to the bench 
plane and it passes in the center of the stop pupil between the L1 and 
L2. It is a 40mm diaphragm on the same mounting of the static screens. 
The 40mm aperture defines the simulated telescope pupil and the F/20 
angle, it is placed in focal plane of both the L1 and L2. The tip-tilt 
of the pupil is corrected in order to be orthogonal to the optical axis 
posing a flat mirror on the diaphragm. The positioning of the tip-tilt 
of the pupil is better than 20arcsec, while the precision in the height 
positioning of the pupil is better than 0.5 mm.

The optical fibers that mimic the star sources can be inserted on a 
large number of slots on a plate. This plate can be moved in z (the 
optical axis of the telescope simulator) using linear stage. The 
telescope simulator re-images a 1:1 copy of the reference object (the 
fiber plate); the 2arcmin FoV corresponds to a circular region of 94 mm 
diameter on the plate where the optical fiber can be arranged in several 
different configurations, to mimic different guide stars constellations.

\begin{figure}[h]
\centerline{\includegraphics[width=11.41cm,height=7.95cm]{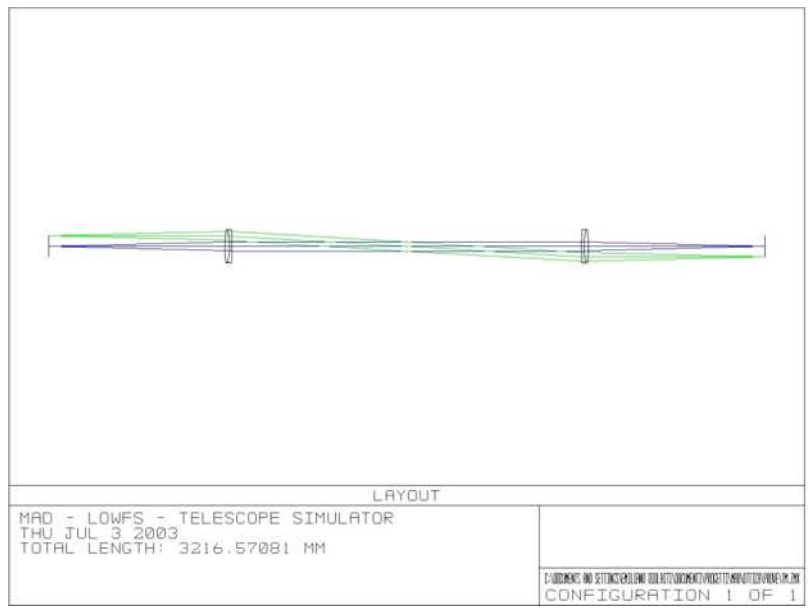}}
\caption{\footnotesize{Optical Layout of the telescope simulator, the 
different rays correspond to different positions of the reference 
optical fibers on the fiber plate.}}\label{fig:2}
\end{figure}

The optical fibers used in the measurements presented in this document 
have a 50$\mu$m core. They do not transmit an identical amount of light, so 
they were selected opportunely, according the test to be performed. 
Regarding the optical quality tests (see section\ref{sec:2.1}) the fiber core dimension does not affect the results 
because the spot dimension on the CCD is produced by the holes size in 
the mask placed at the pupil plane. Furthermore the dimension of an 
ideal source imagined in the LOWFS focal plane by the telescope 
simulator has 180$\mu$m of diameter. However we checked that using fibers 
with different core sizes works such as using different modulation of 
the pyramids and that it changes the calibration factor that transforms 
WF measurements from arbitrary to real length units.

\subsection{Telescope Simulator optical quality}
\label{sec:2.1}
The telescope simulator exhibits a number of potential departures from 
the optical system composing MAD. These are hereby listed and discussed:

\begin{itemize}
\item Field curvature. This is taken into account and corrected for by 
adjusting the fibers positions along the optical axis on the 
fiber-plate, when they are placed off-axis. The defocus measurement is 
made using the pyramid WF sensor embedded in the instrument and it 
achieves usually a quality better than about 40nm.
\end{itemize}

\begin{figure}[h]
\centerline{\includegraphics[width=14.99cm,height=7.58cm]{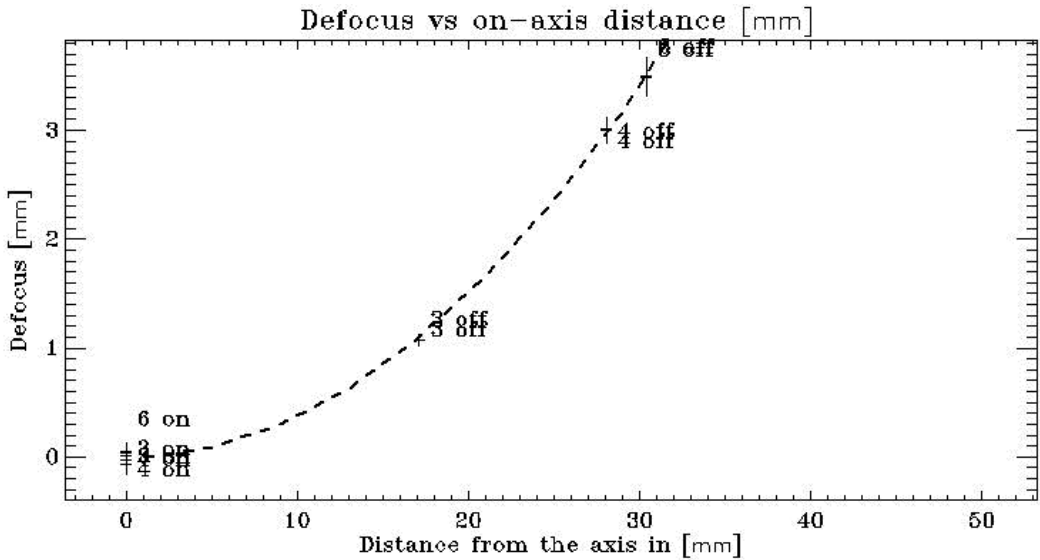}}
\caption{\footnotesize{In this plot a portion of the field curvature 
introduced by the telescope simulator lenses is presented. Using fibers 
more and more off-axis the defocus term is more and more large. Large 
departure from a flat focal plane had been predicted and taken into 
account.}}\label{fig:3}
\end{figure}

\begin{itemize}
\item Telecentricity and final F/20 have been checked during the 
alignment procedure through auto-collimation in the pupil plane and it 
is well within the specifications of the MAD exit relay. According to 
the optical requirements for the input beam of the LOWFS an F/20$\pm$0.2 is 
requested and with a maximum non-telecentricity angle of 1.9arcmin. The 
final focal ratio (15$\times$F/20=F/300) has been checked looking to the 
dimension of the re-imaged pupils on the detectors: we measured 385$\mu$m $\pm 4\mu$m
to be compared with 388$\mu$m (by the optical design). Moreover, considering 
that the star enlargers increase the total focal ratio by a factor 15, 
it is possible to obtain for the telescope simulator an F/20$\pm$0.2. For 
telecentricity has been reached a 1mm precision in the positioning of 
the lens L2 with respect to the pupil mask on the screens holder, which 
corresponds to an error of 50arcsec of non-telecentricity angle.
\item Chromatism is, by Zemax design, very small and negligible with 
respect to the one introduced by the pyramids. Recall that the telescope 
simulator is made up by a couple of achromatic doublets that introduce 
an overall blur of 1/50 of sub-aperture.
\item The residual aberration is mainly astigmatism of third order. This 
is not negligible and depends upon the position on the Field of View. 
However, to perform phase screens analysis a measurement of the static 
aberrations with for each guide stars constellation is performed and 
later subtracted from the WF measurements. This is somehow similar to 
the handling of non-common path aberrations as will be done on the MAD 
experiments, even if in this case we will manage them as reference slope 
with respect to which the MCAO loop will be closed.
\end{itemize}

\section{Calibration of the LOWFS}

The LOWFS has been calibrated in order to translate the differential 
illumination measurements of the 4 metapupils in terms of wavefront 
distortions measured in wavefront displacement units (nanometers).

To achieve this goal we move of a known quantity the fiber-plate along 
the optical axis using a linear stage, which axis was previously aligned 
to the optical axis (defined by the laser beam direction). 

We use the formula:
\begin{equation}
PV = \frac{d^2\delta f}{8f_{0}^2}
\end{equation}

this formula relates the Peak to Valley (\textit{PV}) of the defocus 
aberration to the shift applied to the object on the focal plane along 
the optical axis. In our case \textit{d} is the pupil diaphragm 
diameter (40 mm), $f_{0}$ is the focal length of the first 
lens (793 mm) and $\delta f$ is the amount of shift to be read on the 
linear stage. The validity of this formula was checked using Ray-tracing 
both giving $PV = 90.46$nm. 

Experimentally was found the calibration coefficients to translate 
wave-fronts from arbitrary to nm units.

\section{Pupil image optical quality}

\begin{figure}[h]
\centerline{\includegraphics[width=6.40cm,height=4.78cm]{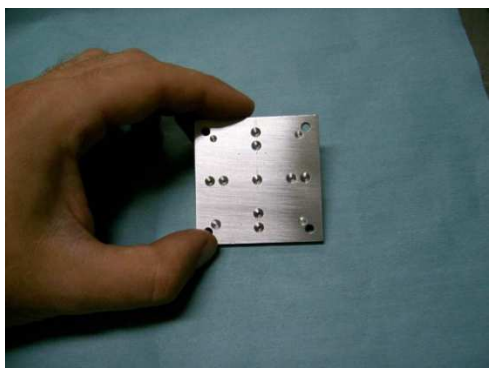}}
\caption{\footnotesize{The picture shows the pupil mask 
with 9 holes described in the text above. Each hole has 0.8mm size.}}\label{fig:4}
\end{figure}

In order to check the optical quality of the WFS objectives we used a 
pupil mask with small holes (Figure~\ref{fig:4}) distributed on 
a cross over the 40mm circular aperture defining the telescope simulator 
pupil. The diameter of each spot has been fixed to 1/50 of the pupil 
size (0.8mm) to fit the dimension of the CCDs pixel size corresponding 
to 1/52 of the pupil diameter. In this way, and in pure geometrical 
approximation, the holes images present a dimension equivalent to one 
pixel, when the mask is placed on the pupil position. In fact one pixel 
is 7.4$\mu$m while the image of the hole should be 7.7$\mu$m. We measured the 
dimension of the spots using only one fiber source first on-axis and 
later off-axis. The image of the mask has been taken for each star 
enlarger (2 lenses and 1 pyramid). The fiber light source emits white 
light centred at 0.4$\mu$m with a large spectrum. In order to avoid 
un-realistic chromatic-effect due especially to the pyramids the white 
light should be modulated in wavelength with a filter as close as 
possible to the final LOWFS spectral characteristics. Such a filter was 
not available and we checked how the measurements fit the theoretical 
previsions.

The fibers illuminator has a bandwidth of 0.25$\mu$m to 0.80$\mu$m with a 
3250Kelvin colour temperature at the maximum intensity. The Electrim 
1000N has a known spectrum response. In this test the two LO objectives 
are imaging an object (the 9 holes mask) positioned at the pupil (the 
ground) position. The pupil of the star enlarger + objective system is 
defined by the dimension of the second lens of the star enlarger (12.7 
mm diameter) however it is not completely illuminated (only about 10 mm 
are effectively used). Considering the LOWFS-MAD central operation 
wavelength is 0.55$\mu$m and the 115.7 mm objective focal length, the $\lambda$/D 
corresponds to a 6.4$\mu$m FWHM. But the image chromatic elongation produced 
by the pyramidic prism overcomes the diffraction effect. In fact the 
divergence angle,$\beta$, of the 4 beams exiting each pyramids depends on the diffraction index 
through the {equation\cite{2003SPIE.4839..164E}}:

\begin{equation}
\beta = \left(n-1\right)\alpha
\end{equation}

where $\alpha$ is the pyramid physical vertex angle (here 1.176 degrees) and \textit{n
} is the refraction index. In particular the pyramids are made of BK7 
glass, which spectral behaviour is presented in Figure~\ref{fig:5}:

\begin{figure}[h]
\centerline{\includegraphics[width=11.11cm,height=6.97cm]{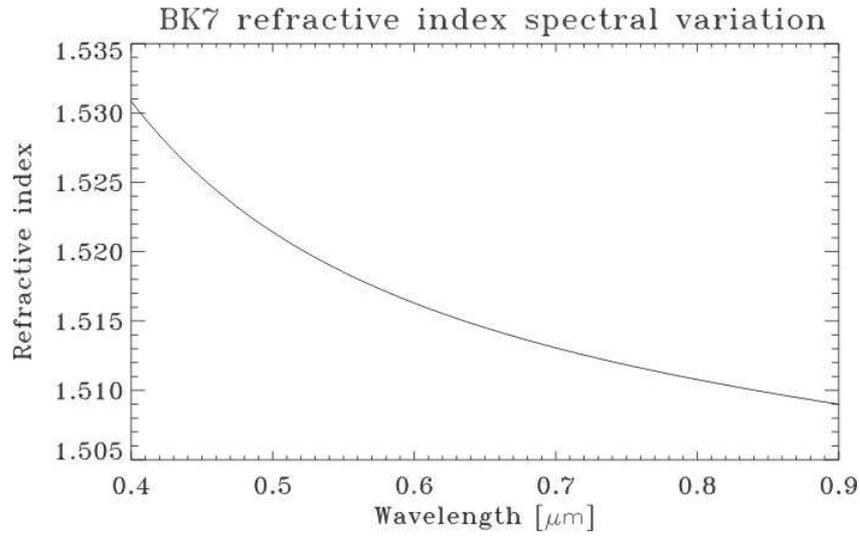}}
\caption{\footnotesize{This picture shows the refractive 
index of the BK7 glass used to manufacture the pyramids with respect to 
the wavelength.}}\label{fig:5}
\end{figure}

\begin{figure}[h]
\centerline{\includegraphics[width=10.56cm,height=7.94cm]{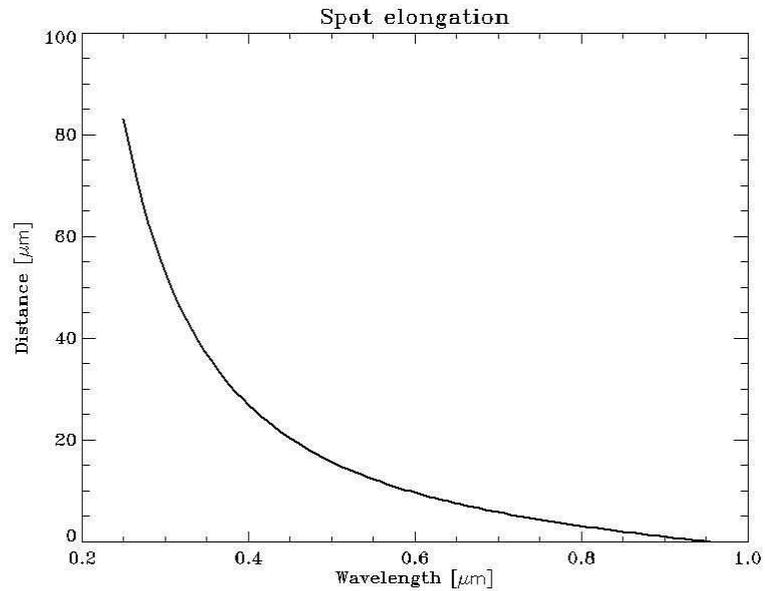}}
\caption{\footnotesize{This plot shows the elongation of 
the spot due to the different spectral components of the light of the 
reference source. In the laboratory case here discussed the spectral 
range is larger. One extreme of the spectral range is fixed to 0.9$\mu$m, 
while the smaller one varies between 0.2$\mu$m and 0.9$\mu$m.}}\label{fig:6}
\end{figure}

\begin{figure}[h]
\centerline{\includegraphics[width=17.07cm,height=5.70cm]{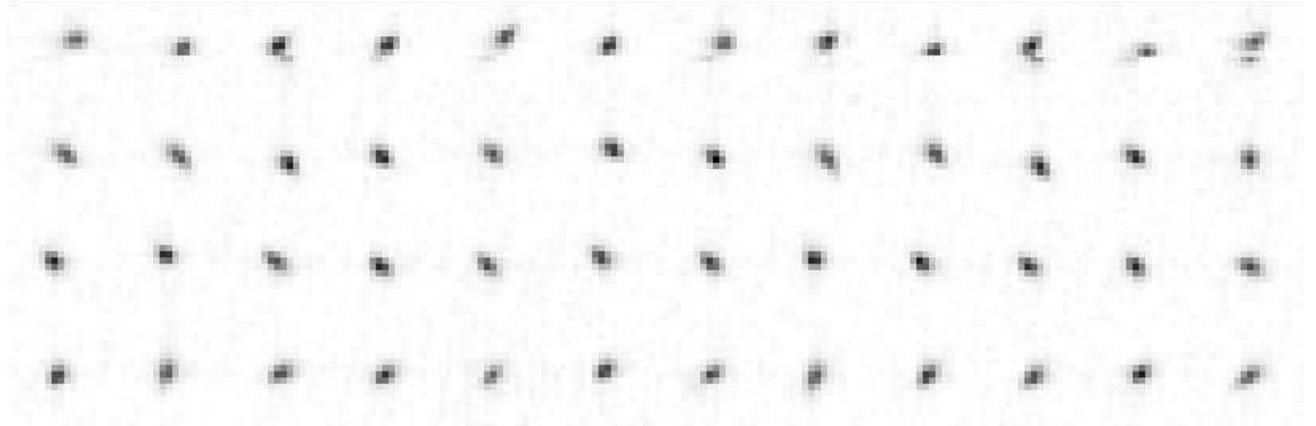}}
\caption{\footnotesize{This picture shows the images of the PSFs generated 
by the high LOWFS1 objective conjugated to the ground plane where the 9 
holes mask has been placed. The 9 holes generate a pretty identical 
image and here only the central one is shown. The 4 lines correspond to 
the 4 quadrants generated by the pyramid faces. The first 8 columns 
refer to different star enlargers looking the same reference fiber 
placed at centre of the 2arcmin field of view. The elongation direction 
is directed toward the 4-pupils barycenter. The last four columns 
present the images of the central hole of 4 different star enlargers but 
in two off-axis positions.}}\label{fig:7}
\end{figure}

This refractive index variation produces a spot elongation,  
larger and larger as the bandwidth used increases (Figure~\ref{fig:6} and Figure~\ref{fig:7}). Moreover the 
elongation directions are defined by the pyramid faces orthogonal 
planes, corresponding to the direction of the 4-pupils barycenter if 
projected on the CCD sensor

\begin{figure}[h]
\centerline{\includegraphics[width=10.07cm,height=7.55cm]{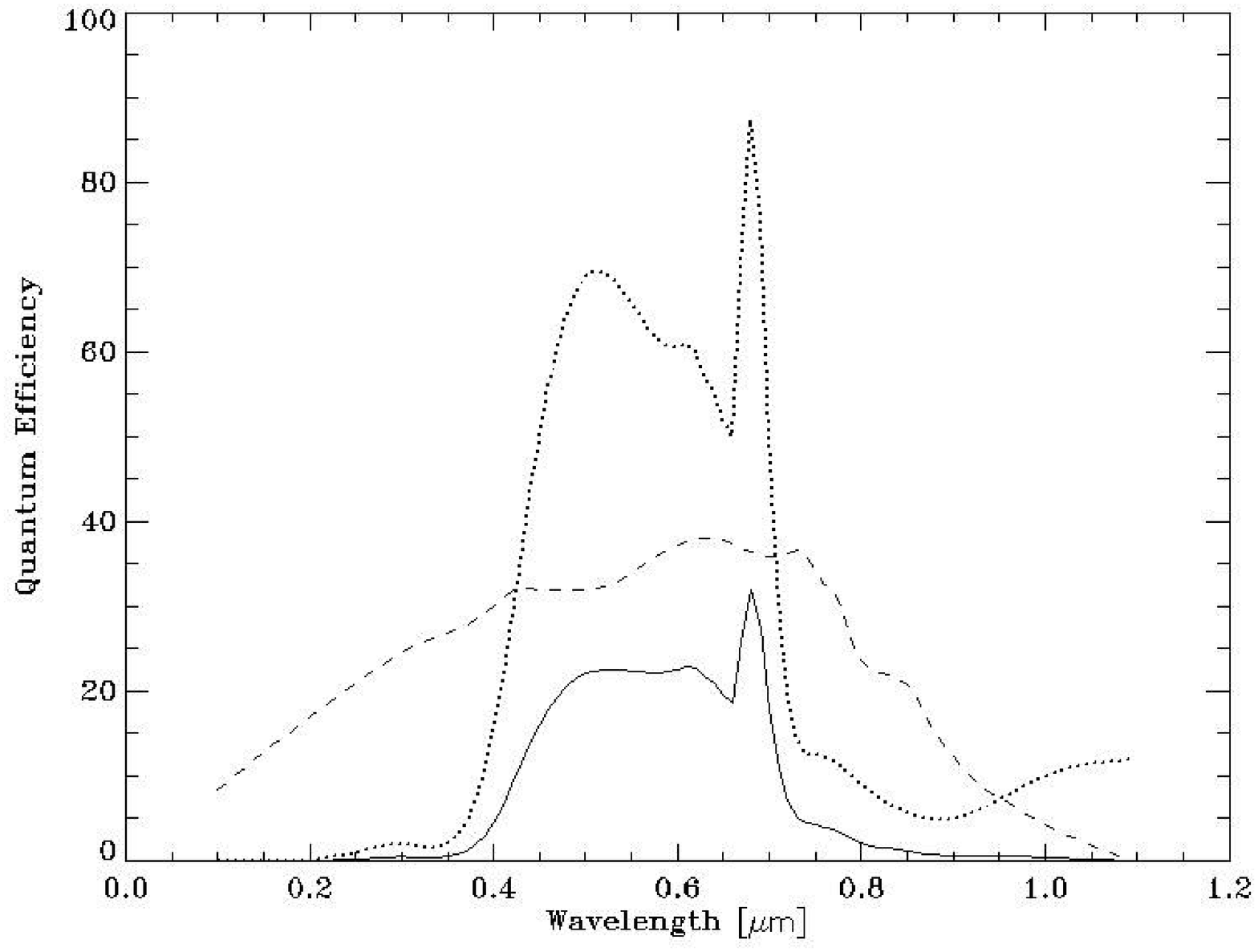}
\includegraphics[width=4.51cm,height=4.51cm]{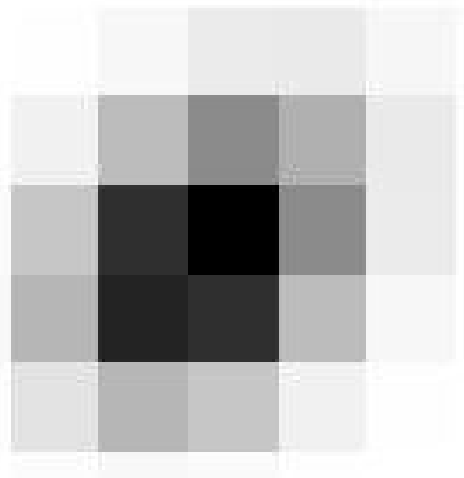}}
\caption{\footnotesize{This plot shows the spectrum and 
the quantum efficiency respectively of the fiber light (dotted line) and 
of the Electrim 1000N CCD (dashed). The solid line represents the 
combination of the two. A 0.4-$\mu$m spectral bandwidth (over 20\% of the 
peak value) in the range 0.4$\mu$m-0.8$\mu$m has been actually used. The picture 
on the right shows the elongation spot simulated by using the quantum 
efficiency curve shown in the left panel. The position of the 
diffraction-limited spot was computed for a set of $\lambda$ (in the range of 
interest), and the image summed according the position found, each one 
intensity-weighted accordingly to the quantum efficiency. In this way 
was found 7.34$\mu$m (considering the spot-rms of the interpolated gaussian) 
or 7.91$\mu$m (on the simulated image (of course with zero noise such as 
Read Out or photon noise). }}\label{fig:8}
\end{figure}

The fiber illuminator emits between 0.25$\mu$m to 0.80$\mu$m. To the wavelength 
range 0.31-0.80 corresponds a 45$\mu$m elongation, equivalent to 6.2 
Electrim pixels. But the emissivity function has not flat spectrum and 
should be multiplied for the Electrim quantum efficiency (QE) response 
function to have a realistic idea of the intensity distribution over the 
elongated spot. In order to check the matching of the measurements with 
the expected results, the latter has been computed with the QE (VS 
lambda) we have in the experimental setup. Please note, however, that 
this is rather flat in the range of interest.

However, we have measured spot elongations with FWHM up to 24.9$\mu$m that 
in a first order analysis gives a maximum elongation of about 50$\mu$m. All 
the point spread functions (PSF) measured have been interpolated with 2 
dimensional Gaussian functions. The fit took into account both 
ellipsoidal shape of the PSF and the axis rotation. 

\begin{figure}[h]
\centerline{\includegraphics[width=9.80cm,height=7.35cm]{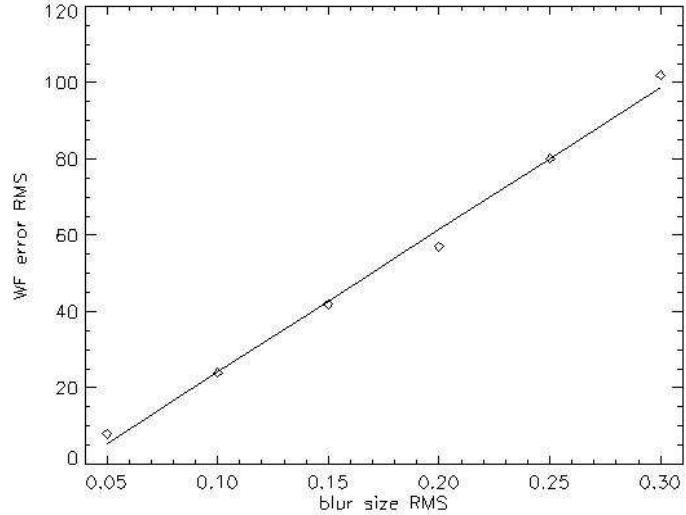}}
\caption{\footnotesize{The picture shows the relationship 
between the rms of WFE (the blur size rms) and the error in the computed 
correction. The values are expressed in nm. This picture was extracted 
from cited {paper\cite{MADLOSDA}}}}\label{fig:9}
\end{figure}

The spot rms computed fitting the measurements data of the PSFs is 
8.04$\mu$m with an rms of 0.58$\mu$m equivalent to 0.167 ground sub-apertures 
(1sub = 48 $\mu$m) equivalent to 52nm WFE according to the Wavefront Error 
Budget analysis performed so {far\cite{MADLOSDA}} (Figure~\ref{fig:9}). In fact the blur of the 
pupil images due to mis-alignment, chromatism, dirty optics, etc.., 
introduce a loss of information in the WF/slope reconstruction, we 
quantified this effect through numerical simulation and it can be 
described such as a linear relation between blur of the pupil image and 
residual phase rms in closed loop operation that exceeds with respect to 
the ideal case with no blur.

In the instrument requirements was given for the chromatism effect a 
57nm rms Wave-Front-Error to be compared to 52nm we have measured in the 
setup used. This error corresponds to an equivalent spot rms of 0.15 
sub-apertures (compared to 0.207). But we saw that elongation effect 
depends on the waveband used. In particular we have to compare the 7.9$\mu$m 
spot rms measure to the elongation spot due to the system effectively 
mounted on the bench. The quantum efficiency of the system illuminator 
and Electrim 1000N was computed (using technical data-sheets) and the 
effective spot elongation simulated. 

\begin{table}[h]
\centering
\begin{tabular}{|l|l|l|l|l|}
\hline
 & Average $[$$\mu$m$]$ & $\sigma$ $[$$\mu$m$]$ & Average $[$sub-aper$]$ & $\sigma$ 
$[$sub-aper$]$ \\
\hline
LOWFS1 ``Ground" & 7.99 & 0.39 & 0.166 & 0.008 \\
\hline
LOWFS2 ``High" & 8.00 & 0.49 & 0.167 & 0.010 \\
\hline
\end{tabular}
\caption{In this table are presented the average 
value of the spot rms values for the different star enlarger on and off- 
axis. The values are given both in $\mu$m and ground sub-apertures. These 
values should be compared to the 7.91$\mu$m theoretically expected.}\label{tab:1}
\end{table}

The final LOWFS system will have a similar chromatic answer and then 
similar elongation spot sizes. More detail about specifications on the 
pyramid vertex angle and measurements can be found in the cited {paper\cite{arcidiacono}}.

\section{Layer Smoothing at different altitudes}

\begin{figure}[h]
\centerline{\includegraphics[width=250.5pt,height=198pt]{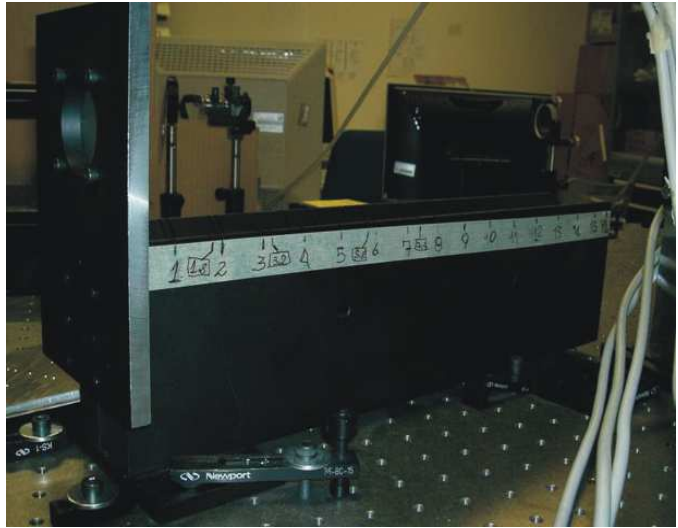}}
\caption{\footnotesize{This picture shows the plastic 
screens holder. In the ``smoothing" test the same screen was moved from 
the position corresponding to 0 km to 15 km.}}\label{fig:10}
\end{figure}

We verified that the LOWFS is able to measure the turbulence at the 
conjugated altitude and in an atmospheric volume close to that plane. 
But the non-conjugated layers are seen more and more smoothed as the 
distance from the conjugation plane increases. To check this statement a 
single plastic screen has been measured with an interferometer and then 
placed in 16 different positions corresponding to altitude between 0 and 
15 km, 1 km spaced, (Figure~\ref{fig:10}). The two objectives had been previously 
conjugated to 0 (LOWFS1) and 9 km (LOWFS2). For this test 3 and 8 
different fiber-reference sources have been used positioned in a 
restricted FoV of 1arcmin for the case with 3 stars and over the full 
2arcmin for the case with 8 stars. The interferometric measurements of 
the plastic screen aberration refer only to a 50.8 mm pupil, to be 
compared to the 66.2mm size of the metapupil at 9 km. In the comparison 
we had not considered the outer region of the LOWFS measurements. For 
both WFS a couple of 4 meta-pupils image with slightly different 
tip-tilt have been taken (introduced de-centring the pyramid optical 
axis using the XY linear stages to simulate a tip-tilt modulation).

A posteriori the static WFs have been removed from the phase 
measurements. Results and comments are in the picture captions.

\begin{figure}[h]
\centerline{\includegraphics[width=13cm]{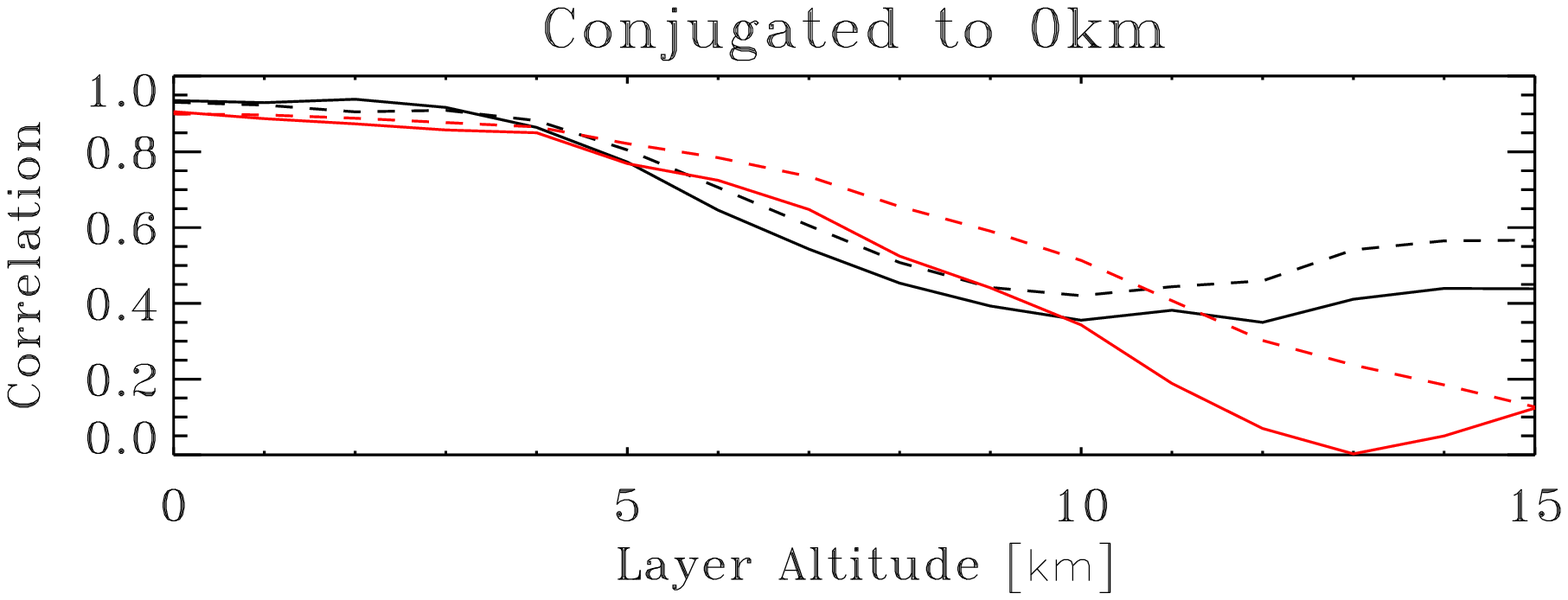}}
\centerline{\includegraphics[width=13cm]{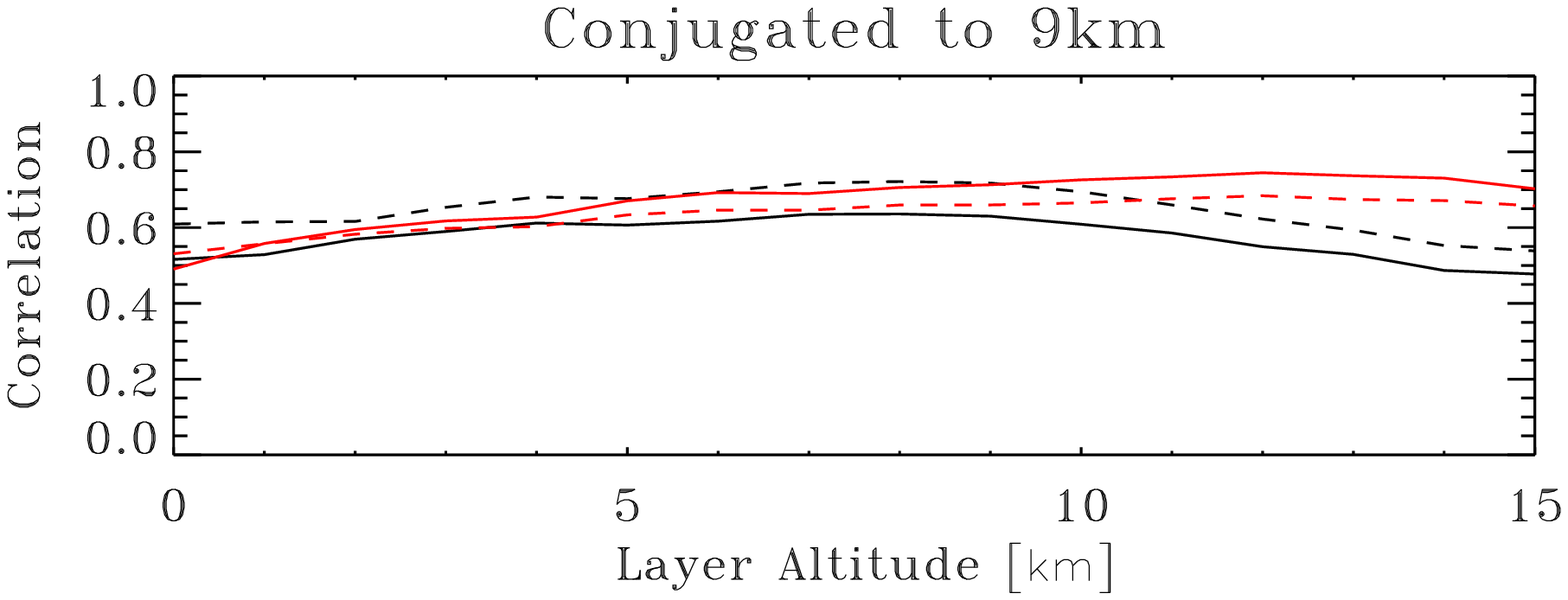}}
\caption{\footnotesize{The plots above show the correlation values between 
the WF measured both with the LO WFS and the interferometer. The 
correlation increases more and more the screen is close to the 
conjugation plane. To neglect high order errors due to processing 
procedures in the interferometric and LOWFS measurements and to the 
different pixel-size the interpolation with the firsts 50 Zernike 
polynomials have been computed and interpolated WF correlated. Dashed 
lines present data without tip-tilt modulation, solid with it. The 
red/grey case refer to the case with only 3 stars in central 1 arcmin, 
the black ones to the 8 stars. Using more stars helps to ``see" better 
high conjugation planes. }}\label{fig:11}
\end{figure}

\begin{figure}[h]
\centerline{\includegraphics[width=9.47cm,height=4.58cm]{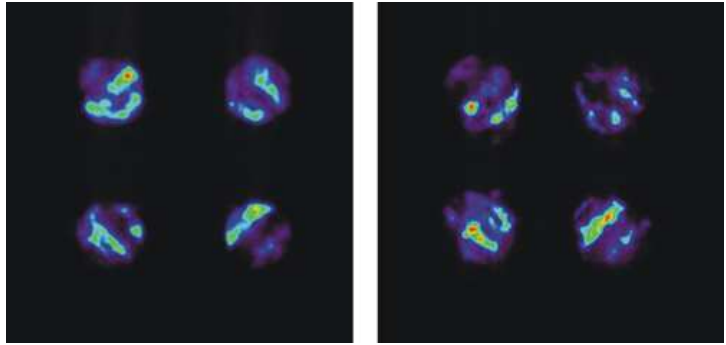}}
\caption{\footnotesize{These pictures show: on the left 
the image of the 4 pupils illumination measurements with ground 
conjugated WFS; on the right the same screen seen by the high conjugated 
(9 km) WFS.}}\label{fig:12}
\end{figure}

\section{Scanning the atmosphere}

In this test we verified how the ground and high WFS measurements are 
still correlated when screen at different altitudes are inserted. In 
particular we first place a screen at 0km, and then we added a screen at 
9 km and finally at 4km. Then we removed, in this order, the 4km and the 
0 km screen. For each of this configuration ground and high conjugation 
plane (9 km) have been retrieved.

\begin{figure}[h]
\centerline{\includegraphics[width=13.58cm,height=7.79cm]{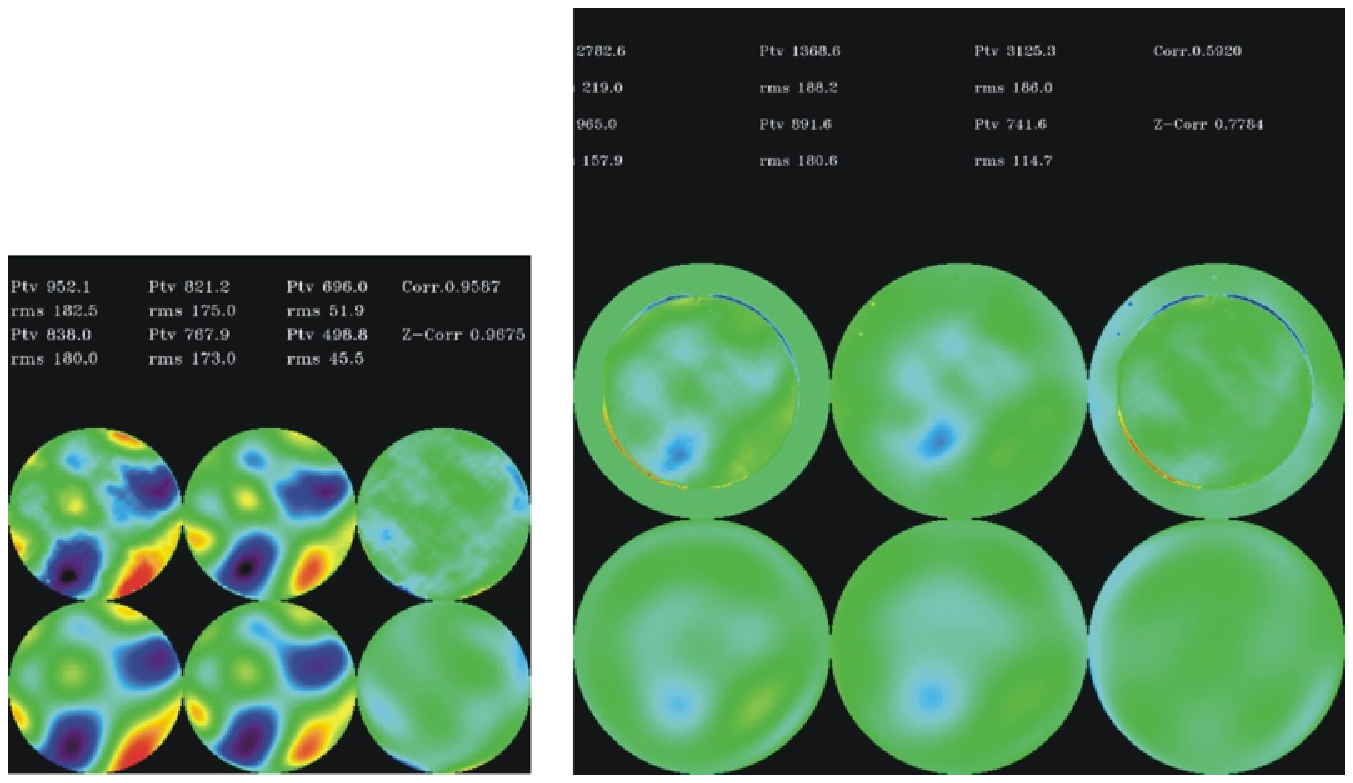}}
\caption{\footnotesize{These pictures show WF 
measurements for: on the left the ground WFS observing the screen at 
0km; on the right the high WFS ones for the screen at 9km. From the left 
to the right (top row) the Interferometer WF data, the LO WFS WF and the 
relative difference, on the bottom the same WF but interpolated with the 
firsts 50 Zernike polynomials. The values written on the top are in nm 
with on the right the correlation values; on the bottom the Zernike 
interpolated WF values. (``Ptv" is WF Peak to Valley, ``rms" the standard 
deviation over the metapupils). Five stars over the two arcmin FoV have 
been used and static aberration removed numerically.}}\label{fig:13}
\end{figure}

A posteriori the static WFs have been removed from the phase 
measurements. Results and comments are presented in the captions of Figure~\ref{fig:12} and~\ref{fig:13}.

\section{Conclusions}

In this paper we show rapidly the more interesting topic related to the 
alignment of the LOWFS for MAD. We described how this had been reached 
and a few indicator of the quality achieved, such as the pupil optical 
quality. Finally we presented several results to show the ability of the 
LOWFS to measure WF both in classical Adaptive Optics and MCAO modes 
comparing plastic screen WF measurement to the ones performed by an 
interferometer. This comparison gives very high correlations especially 
with low strength turbulence, a behavior already predicted by pyramids 
WFS models. In fact it is known how pyramid WF gives best performance in 
closed loop operation rather than open loop one (such as the cases we 
had in experimental setup without deformable mirrors).

\begin{figure}[h]
\centerline{\includegraphics[width=10.5cm]{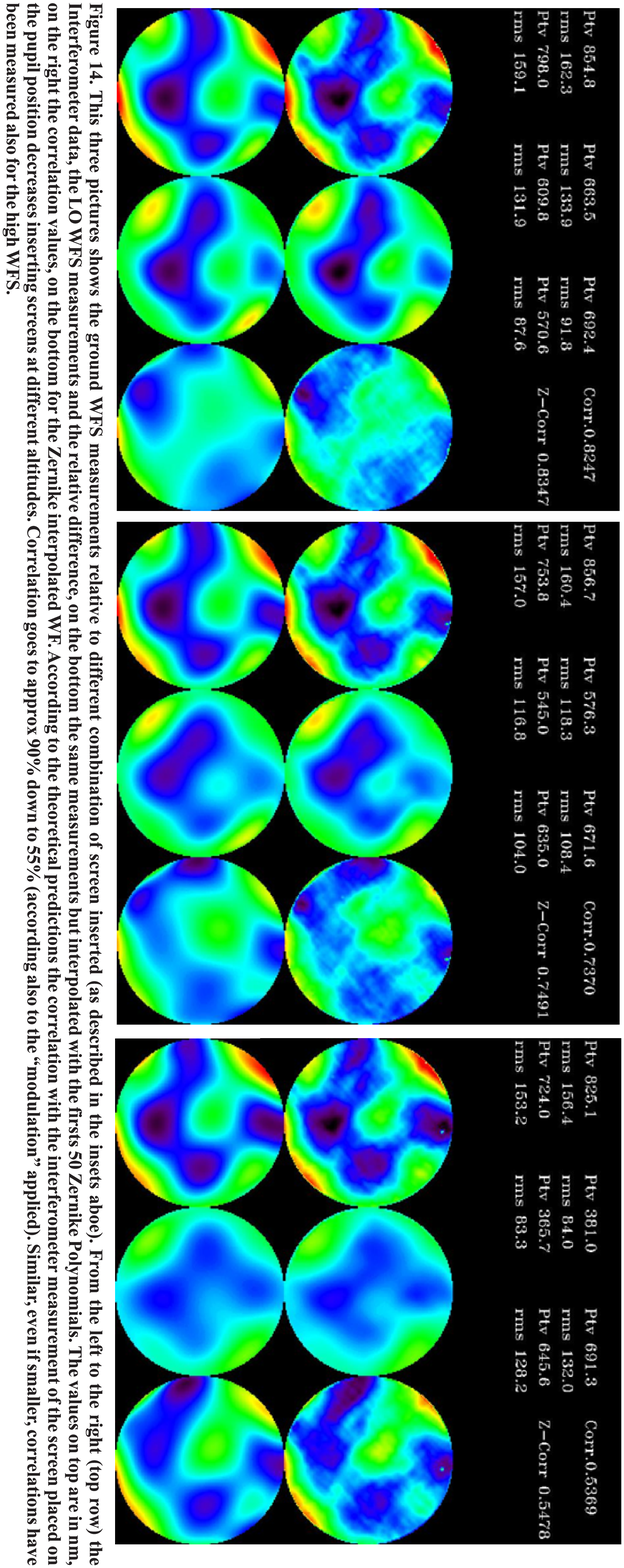}}\label{fig:14}
\end{figure}

\bibliography{spie}   %>>>> bibliography data in report.bib
\bibliographystyle{spiebib}   %>>>> makes bibtex use spiebib.bst

\end{document}